# Performance evaluation of Secure DSRC (S-DSRC) protocol for Vehicular Adhoc Networks

**Surender kumar[1] Dr. Vikram Singh[2]**

**Abstract:** Vehicular adhoc networks (VANETs) provides communication between the vehicles like V2V and also provides communication between the Vehicles and the Road side Units. Recently, VANETs have greater scope in establishing interconnectivity between the vehicles on road and off road. These networks ensure the timely report of accidents happening in highways, alerting the nearest hospital, insurance claims if the vehicle is breakdown, etc. Vehicles often travels in highways lack the network connectivity, which then can be ensured through many technology implementations like 4G, 5G, LTE networks and protocols like IEEE 802.11p and DSRC were helpful in achieving the effectiveness of vehicular networks. In this paper, we have proposed a protocol Secure Dedicated Short-Range Communication (S-DSRC) protocol that can establish connectivity between the vehicles in a secure way. There will be a secret key, private kay and public key exchange between the vehicles so that the safety message can be sent to vehicles in a secure way. Our results show that S-DSRC protocol outperforms the existing DSRC protocol for a medium sized network.

## 1. Introduction

Dedicated short-range communications (DSRC) is one of the key protocols that is helpful for the intelligent transportation systems (ITS). In many countries, ITS is standardized and DSRC was the default communication protocol for safety messages to be sent between the vehicles to vehicles (V2V) or vehicle to infrastructure (V2X). The main objective of DSRC protocol is to deliver the basic safety messages (BSM) reliably to the vehicles concerned during collision, accidents, etc.

DSRC protocol is standardized to work under the frequency range of 5.9GHz along with the WAVE 1609 standard as mentioned by Maddio, S., et al., in their work. Due to the size of vehicle density and speed of the vehicles, the DSRC performance analysis is highly complicated.

Figure 1 shows the DSRC protocol stack along with the 1609 WAVE standard. In this protocol stack, a new module called the transport security agent is attached along with TCP protocol which makes this protocol as a S-DSRC protocol. The safety message packets which are been sent to vehicles should be encrypted before its being sent. Hence while the transport layer is active, the security protocol is added at the transport layer. Also this figure mentions other blocks namely physical layer with IEEE 802.11p, Mac Layer with 1609.4 and it also has the inbuilt security of 1609.2 which is the default DSRC security layer. This protocol stack helps both safety and non-safety DSRC applications.

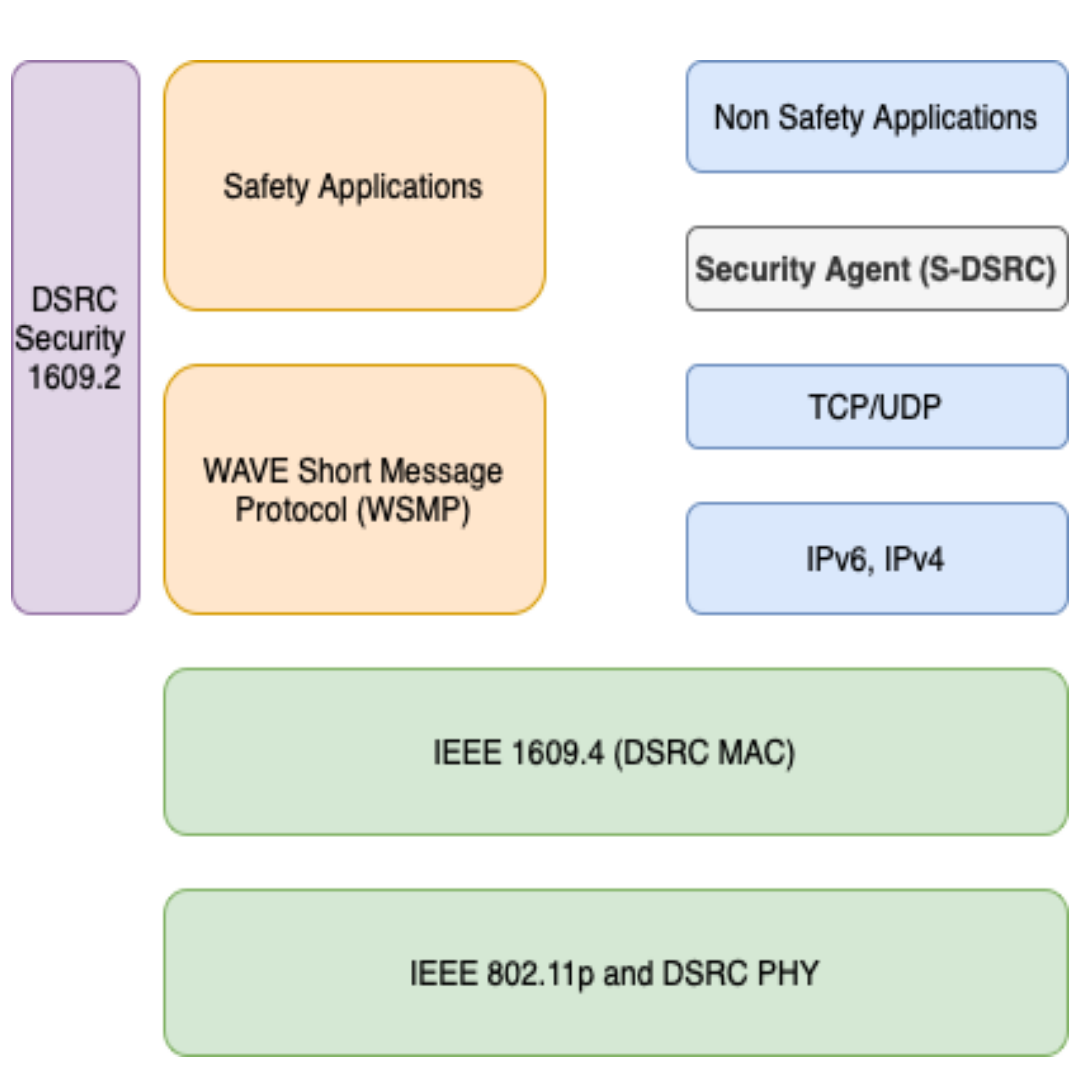

**Fig 1.** S-DSRC Protocol Stack

## 2. Related Works

DSRC protocol is mainly for sending safety critical messages or basic safety messages to the vehicles that follow the standards of intelligent transportation systems (ITS). Some of the works carried out by many researchers over these years have proposed, implemented, suggested

*Department of computer science and engineering*
*Choudhary Devi Lal University*
*Sirsa, Haryana India*
*skgill.ugcnet@gmail.com*
*Professor*
*Department of computer science and engineering*
*Choudhary Devi Lal University*
*Sirsa, Haryana India*

many additions, improvisation of DSRC protocols on security, clustering, energy management, usability, etc. Some of the literature is mentioned below.

Zhang, M et al., proposed a medium access Control protocol based on the DSRC protocol. This protocol is well suited for the purpose of basic safety messages which is been sent to nearby vehicles during a collision or any accidents. This protocol is designed in a such a way that it obeys the IEEE 802.11p protocol as well. Maddio, S., et al., implemented a reconfigurable leakage canceller for the carrier frequencies in the range of 5.8 GHz. This process will delete or suppress the leaked signal emanating while using the DSRC. Liang, C. N et al proposed a test bed that emulates the DSRC protocol using a 4-machine system where 3 machines runs the WAVE boxes and one personal computer. Among that two devices act as road side units and one unit resides in a car. This paper also uses a jammer that can be interfered by multiple cars. Also, this paper handles the packets that were under higher channel capacity.

Kumar, T. P., & Krishna, P. V. and suggested a model that is solved using reinforcement learning which can optimize the power usage in internet of things and internet of vehicles. This paper identifies the power profile of various devices in the system and based on the power profile, the system is modeled using the semi Markov decision process (SMDP) and solved using the reinforcement learning. Schäufele, B et al., proposed a method for DSRC based parking system where the automated parking is handled with minimal efforts. This system uses a low latency due to the short-range communications between the cars to infrastructure and Car to car connectivity. This paper also developed and used a session-based state machine protocol for autonomous driving of vehicles. Wu, S. H et al., implemented and suggested a new energy efficient protocol that can minimize the energy consumption of the DSRC protocol to 44%. This protocol divides the vehicles into many clusters and within each cluster, there will be a different wake-up/sleep schedule so that the average energy consumption will be minimsed. This protocol is based on the IEEE 802.11 power saving mechanism which is then integrated in to DSRC protocol for energy optimisation.

Ajami, A. K., & Artail, H proposed a new method to make use of using Wi Fi to operate under the frequency range of 5.9GHz as the same frequency is fixed for intelligent transportation system namely the DSRC protocol as well. This paper handles and suppress the interferences arising out of the usage of DSRC and Wi Fi in the 5.9GHz frequency range. They use a technique called stochastic geometry modelling. Cseh, C suggested the architecture of DSRC protocol and its layered architecture in this paper. This is one of the oldest literatures where it talks about the architecture implementation of the first version of dedicated Short-range communication protocol. Ng, H. H et al., suggested and implemented a new system that can send information like traffic updates, traffic lights details, GNSS positioning information, vehicle probe data to the autonomous vehicles. The name of the system is BESAFE which allows and obeys the international standard of DSRC protocol and its frequency. Since the system is useful only for the autonomous vehicles, the system has an onboard encryption standard, so that the messages are encrypted before being sent to the vehicles.

Su, H., & Zhang, X proposed a medium access control protocol that can send safety messages from cluster to cluster. They use content free and contention-based protocol within clusters and between cluster heads respectively. Each cluster heads can relay the safety message in real time to other cluster heads and the heads in turn send those messages to the cluster vehicles. The relay of these messages can be sent to both real time traffic and non-real time traffic as well. Kim, S., & Kim, B. J. proposed DSRC protocol to transmit basic safety messages safely to the vehicles which are prone to meet with accidents. They use an algorithm to find the crash risk vehicles based on the distances and the speed of the vehicles between them. Ahmed, M. S et al., demonstrated the use of DSRC protocol in an android application where the real time vehicle movement is captured in a real-world map. They use Bluetooth communication between the antenna and mobile. Many DSRC units were used and it captured the movement of all vehicles that is enabled with a DSRC kit in it.

Anjum, S. S et al., designed a radio frequency-based sensor network for energy harvesting and hence the optimal energy is being used while transmitting and receiving. This paper uses a model that uses reward allocation when the states transition happens. This work is extended to handle optimal energy management for Internet of Vehicles as well. Hassan, M. I et al., suggested a mechanism to avoid the retransmission of packets when sending the safety messages using IEEE 802.11p protocol between vehicle to vehicle communication. They use a technique called blind retransmission which improves the packet delivery ratio without compromising the degradation of the safety messages. Mir, Z. H et al., have proposed a vehicular network that integrates the dedicated short-range communication and cellular network that can establish connectivity between the vehicles and the infrastructure seamlessly. They use the radio access technology to measure the QoS guaranties and dynamic communication between the V2V and V2X systems. They also measure the network metrics like packet delivery ratio, latency and throughput.

Ikawa, M et al., proposes the DSRC protocol and its architecture for the first time in Japan and DSRC being used as a local communication platform. This protocol is then customized to use it as a push service as some of the aspects of payment, parking access, etc. will be pushed to customers. Bae, J. K et al., did a thorough study on experimenting the WIFI based DSRC protocol for handling the traffic congestion and accidents in vehicular networks. This system had well established connectivity between Vehicle to vehicle (V2V) and Vehicle to infrastructure (V2X). They also use drones for handling the traffic congestions. Their experimentation is done at urban road, tunnels, signal intersections, etc. Ho, K. Y et al., implemented and simulated wireless access in vehicular environment with a dedicated short-range communication in a on-road simulation. The performance of the system is evaluated using a protocol called WSMP (WAVE short message protocol).

Mak, T. K et al., suggested a method for vehicular infotainment and safety in two different channels using a new hybrid-based access point. The name of the access point is Coordinating access point (CAP) which is based on the IEEE 802.11 distributed coordination function and point coordination function. There are two service channels, one channel will handle the communication between the vehicles infotainment system and the second channel handles the critical safety tasks between the vehicular communications. Zhang, M et al., proposed an analytical evaluation of increasing the packet delivery ratio of the vehicular network through a new hybrid MAC protocol. The hybrid mac protocol handles the packets and the transmission in one dissemination period so that the delivery latency is too small.

## 3. Preliminaries and Definitions

Existing security protocol is described in this chapter and introduced the S-DSRC protocol for Vehicular Adhoc Networks. The Secure DSRC protocol enabled with a private key, public key and a secret key that can enable the Vehicles to exchange information between other vehicles on road. A cloud model is also being used to send the information to the cloud so that the network would be capable of handling the messages whenever there is any accident or any breakdown in the roads. This chapter introduces the definitions of various security schemes that are already implemented in various applications.

**General Assumptions:**

The following variables are used in this paper.

**Table 1.** Variables and assumptions used

| Variables | Definition |
|---|---|
| P1, P2 | Cyclic Groups |
| ha1, ha2, ha3 | Hash functions |
| m | Text message |
| m` | Cipher text |
| N | Whole number for Modulo |
| e | Encryption key |
| V | Number of Vehicles on road that is part of a network |
| $G_k$ | Secret key |
| $P_k$ | Public Key |
| $A_k$ | Attributes of vehicles. |
| $AttrID_i$ | Attribute Key |
| a | Public key |

**Bilinear maps assumptions:**

Let P1 and P2 be the two groups of order o which is a prime number and x is the random number generator of P1, then we can get a map with the following model

$$P1 * P1 \rightarrow P2 \quad (1)$$

Which is a bilinear map which can hold the properties of

- computability ($e(u, v)$),
- non-degeneracy ($e(u, v) \neq 1$) and
- bi-linearity (*For all* $a, b \in Z_p{}^*$ *and* $u, v \in \mathbb{G}_1$ *then it hold* $e(u^a, v^b) = e(u, v)^{ab} = e(u^b, v^a)$.)

**RSA Assumptions:**

It is tedious to compute *m* from given cipher text *m'*, encryption key *e* and a modulus *N* of unknown factorization such that $m = m^e \bmod N$. It is hard to find any pair of *(m, e)* such that $m^e$ is congruent to the given cipher text *m'* with respect to a modulus *N* of unknown factorization.

**Diffie-Hellman assumptions:**

Let $a, b, c$ *and* $x \in Z_p{}^*$ and chosen randomly, if *g* be a generator of group $\mathbb{G}_1$ then the Decisional Bilinear Diffie-Hellman (DBDH) assumption can be define as, there is no such probabilistic polynomial time algorithm (PPT) that can distinguish $e(g, g)^{abc}$ from $e(g, g)^z$ given $A = g^a$, $B = g^b$ , $C = g^c$, with more than negligible advantage.

## 4. Proposed method and algorithm

In this paper, we design a Secure DSRC protocol that can exchange information from Vehicle to Infrastructure (V2I) and Vehicle to Vehicle Communication (V2V) in a secure way that can protect from the attacks like Black hole and Sybil attacks which are common in vehicular networks. Figure 3. Depicts a typical diagram that shows the vehicular networks and the cloud environment where the secure messages were encrypted and decrypted. The S-DSRC works in these following 4 phases.

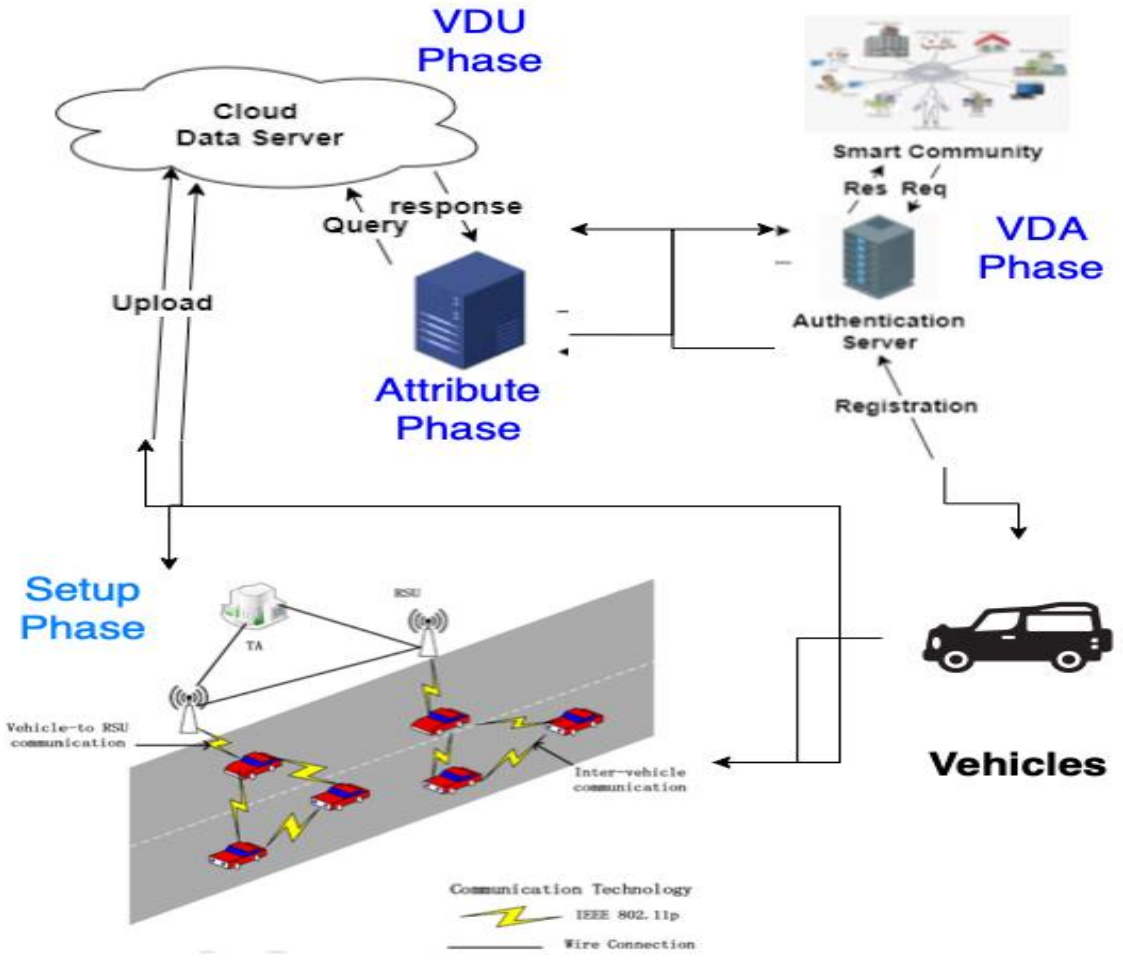

**Fig 3 –** System model with various phases.

***Setup (SP) phase*****:** This phase usually handled by the vehicles that are part of a network on road. It takes input λ as security parameter and generates a secret key $G_k$. The vehicles selects a bilinear pair e (P1 x P1 -> P2) as per (1) and it generated three hash functions namely ha1, ha2, and ha3. The vehicles calculates the public key $P_k$ and derived attribute key $D_i = Q_i * G_k$ for each $A_i$. Vehicles shares the $D_i$ only with other vehicles that is part of network, made $P_k$ public and kept $G_k$ secret.

- Selects a bilinear pair $e(P_1 \times P_1) \rightarrow P_2$ such that $P_1$ is additive group with generator $g_1$ and $P_2$ is a multiplicative group.
- Choose ha1, ha2, and ha3 as three different one way hash functions where $ha_2: \{0, 1\}^* \longrightarrow G_1$ and $ha_3: G_1 \rightarrow \{0,1\}$
- The vehicle selects a global secret key $Gk \in Z_q^*$, where q is a prime number and derives public key *a*
- Based on number of data attribute the vehicles selects Attribute Authority $(A_k)$ and assign responsibility to manage corresponding data attribute $(AttrID_i)$ i.e. $A_i$ is whole and sole responsible for the $AttrID_i$
- Calculate $Q_i = ha2(AttrID_i)$ and $D_i = Gk * Q_i$ for each attributes of health information record.
- Calculate user key $(ask_i)$ for each $AttrID_i$ as $ask_i$

***Attributor (AA) phase:*** In this phase, the algorithm is executed by the vehicles to add new attribute. For each new $A_k$ of vehicles, *an* unique number namely AttrID is created and which is calculated with (AttrID * $G_k$). The vehicles share *this value* with valid other vehicles through a secure channel. Each attribute authority generates public-private key pair as:

- The Ai selects two large prime number *p* and *q*
- Calculate $M_i = p_i * q_i$, $A_i$ Selects $e_i$ and calculate $d_i$ in such a way that $e_i * d_i \equiv 1 \ (mod\ \Phi(M_i))$
- The vehicle with the attribute id Ai publish $e_i$ as its public key

**Vehicular Data Upload (VDU) phase:** The vehicular data upload to the cloud is handled either the vehicles or by the Road side Units. For all $\{AttrID_i\}_{i=1...n}$, if the vehicles wish to upload $AttrID_i$ then have to encrypt it with the encryption function *Enc( )* and securely associate the encryption key with ciphertext as $CT_i = [m_i, S_i, T_i]$, otherwise choose three random number and upload as pseudo cipertext without any computation, as $CT_i = [rand_1, rand_2, rand_3]$. We assume, the system uses the secure encryption algorithm *Enc( )*, which takes the $r_i$ as encryption key, randomly chosen for $AttrID_i$. Vehicles then creates the access policy APi for each $AttrID_i$. Finally the vehicles upload the cipher text and access policy along with the hash value of its identity.

**VD Access (VDA) phase :** This is the last phase to access the vehicular data from the cloud by the RSU or by other vehicles. The attributes of the vehicles will be accessed through the $AttrID_i$ After getting the response from the other vehicles, the attribute key will undergo the decryption process to retrieve by other vehicles. Also for the data access by the Road side units, apart from the attribute key, the secret key also to be supplied, since the RSUs are not part of the network. So the secret key should also be maintained in the cloud.

- Based on the assigned responsibility, RSU and other vehicles are allowed to access maximum k attributes of stored record from the cloud, where $0 \leq k \leq n$
- When data user request to access $AttrID_i$, the vehicle checks the user credential, if it fail to authenticate, return
- Otherwise, the vehicle provide $S_j$ to user and send data access notification with user identity to original vehicle which uploaded the attribute to the cloud.

**Road Side Unit Access Request**

- RSU Data request compute $y_j = H_1(ky_j, atr_j)$ and $V_j = y_j * \text{AttrID}_i$ where $ky_j \in Z_q$
- Select n random number for each attributes $\text{AttrID}_i$.
- If data user wish to access $atr_j$ then computes $\eta_j = rn_j^{ej} * S_j$ otherwise $\eta_j = rn_j$
- Selects re-encryption and decryption key as $e_r * d_r \equiv 1 \ (mod\ \Phi(N_r))$
- Sends the PHI access request as $RQ = [\,[RQ_1, RQ_2, ..., RQ_n], e_r, idh]$

These phases were then verified for experimentation with S-DSRC protocol that is running with a private cloud to store the information and network simulator 2 was used to enable the DSRC protocol. These algorithms ensure confidentiality, anonymity, secure data access and prevents insider attacks.

## 5. Results and experimental analysis

To experiment the S-DSRC protocol, we have used network simulator 2 (ns-2.35) and IEEE 1609.2 WAVE standard protocol and designed the four phases within the DSRC module. The table 2 lists the various experimental parameters used in our simulation.

**Table 2 –** Parameters used

| Name of parameter | Value of Parameter |
|---|---|
| Number of Vehicles | 20 to 200 vehicles |
| Vehicle density (Vehicles/km) | 20 to 200 / km |
| Data Rate | 10Mbps, 20 Mbps |
| Packet length | 512 bytes |
| Protocol used | DSRC |
| Security Patch | S-DSRC |
| Range | 2 to 4 kms |
| Cloud | Private Cloud based on Xen Citrix. |
| Road Traffic | SUMO (Simulation of Urban Mobility) |
| MAC Layer | IEEE 802.11p |
| Transport layer | BSM under TCP and UDP |

In our simulation, we have used ns-2.35 with the DSRC patch and we have added the S-DSRC as per the model designed in this work. The road traffic is enabled using SUMO tool (Simulation of Urban Mobility) where a real road is taken through Open Street Maps (OSM) and also a user defined road structure is used for simulating S-DSRC protocol and results were compared against each other. Figure 4 and Figure 5 shows the two-road structure used in this simulation. In Figure 4, user defined road structure is taken where all the roads are of two-way roads with a right-hand traffic is enabled. The vehicles running in the road can talk to each other or exchange BSM to each other since the roads were designed in such a way that the distance of the roads is longer than the transmission distance. Figure 5 shows the open street map of a place in Haryana and a radius of at least 2 kms is taken for simulation.

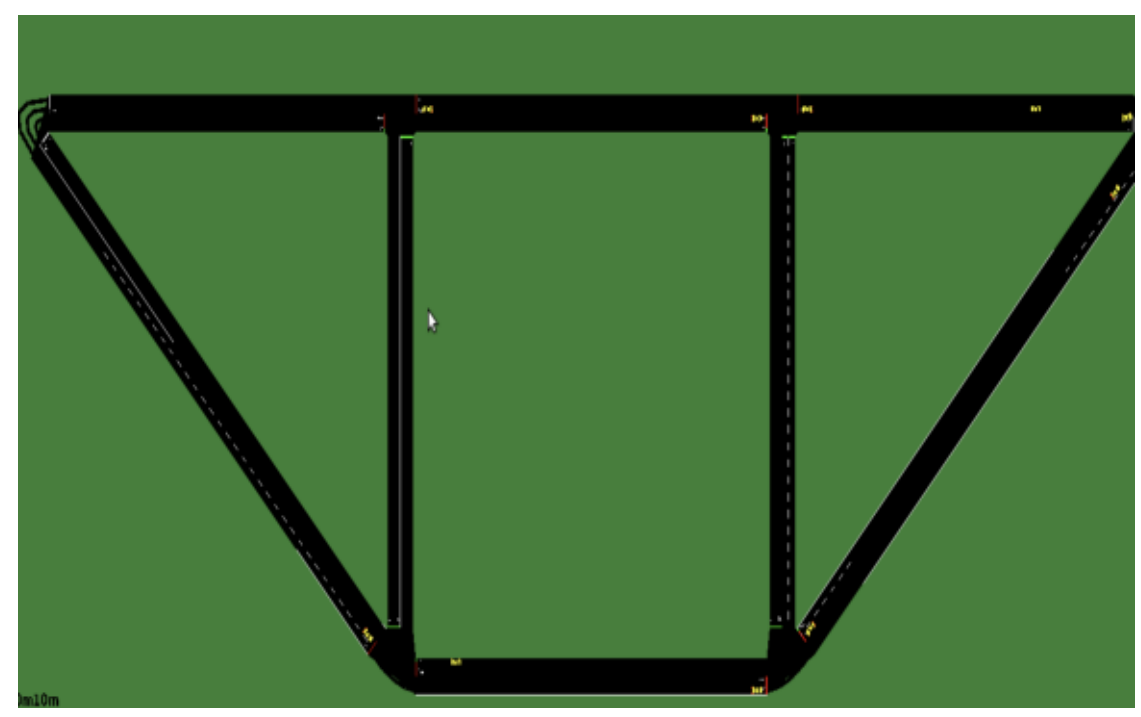

**Fig 4 –** User Defined road structure

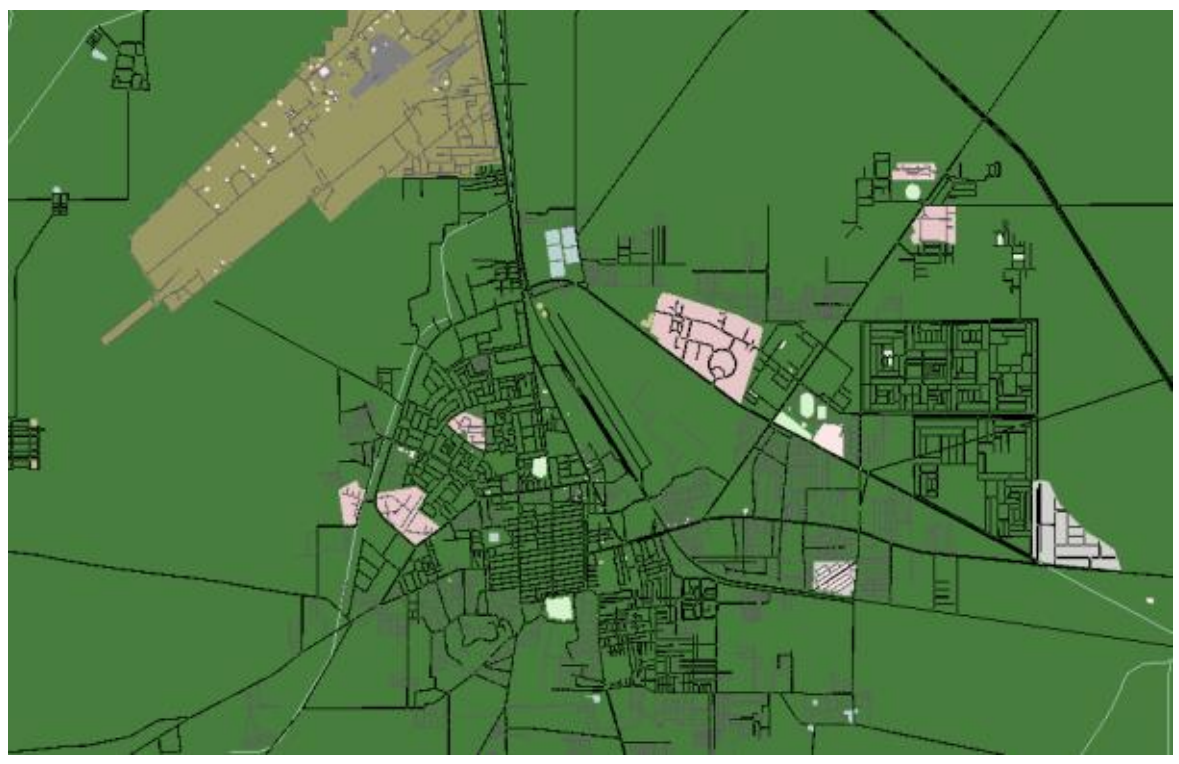

**Fig 5 –** Open Street Map of a location in Haryana (SIRSA)

In our experimentation, we have carried out the experiments with and without security patch in the DSRC protocol. The metrics like Packet Delivery Ratio (PDR) is too crucial as the vehicles are moving at a higher speed, hence the PDR might be affected as the packets need to be encrypted before they are being forwarded. Throughput, delay (message delay and end to end delay), etc. were also calculated.

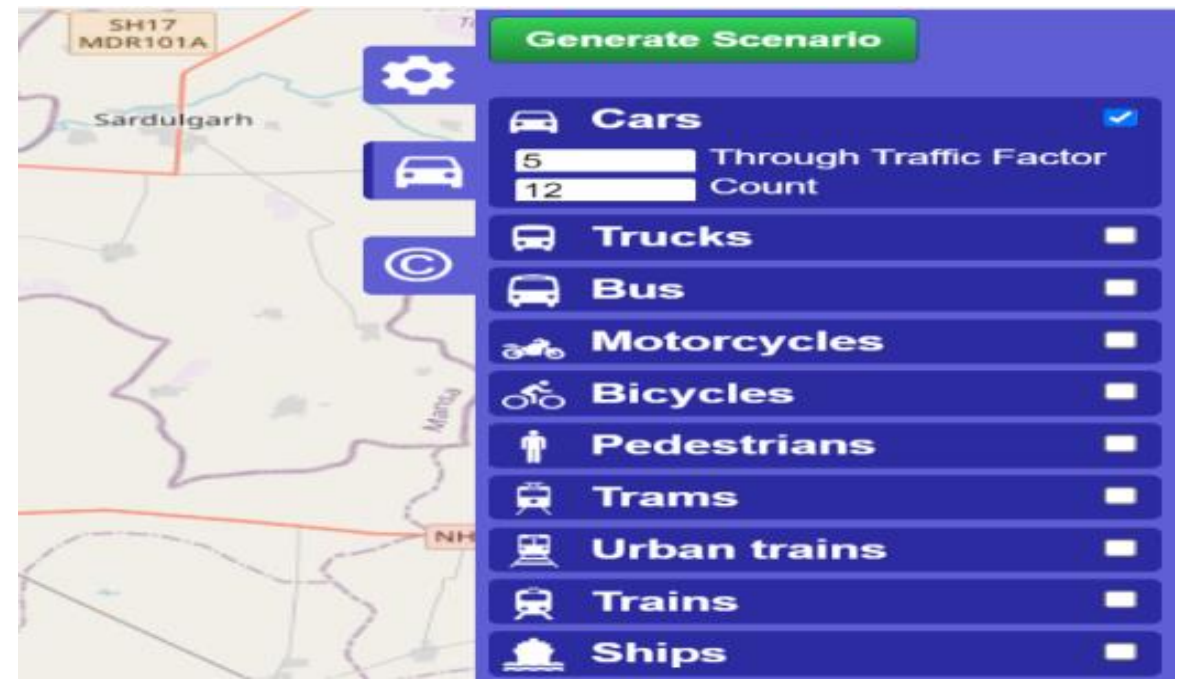

**Fig 6 –** Traffic Factor

All these results were carried based on the number of vehicles assigned per km or vehicle density. As indicated in Table 2, the vehicles density is calculated based on the number of vehicles floated per km in the SUMO through the traffic factor variable as shown in the Fig 6. For this experimentation, only cars were taken into consideration as the research is on Basic safety message transmission between multiple vehicles and our research is mainly on the cars. But the scope of this research can be extended for other vehicles as well like buses, trucks, etc. The experimental results were plotted with DSRC and Secure DSRC protocol under evaluation with 10Mbps as bandwidth and 512 bytes as the packet size. Since the vehicles moves at a higher speed, if the packet size is too large, there could be a chance of huge packet loss. Hence to avoid that, in this work, a maximum of only 512 bytes is used.

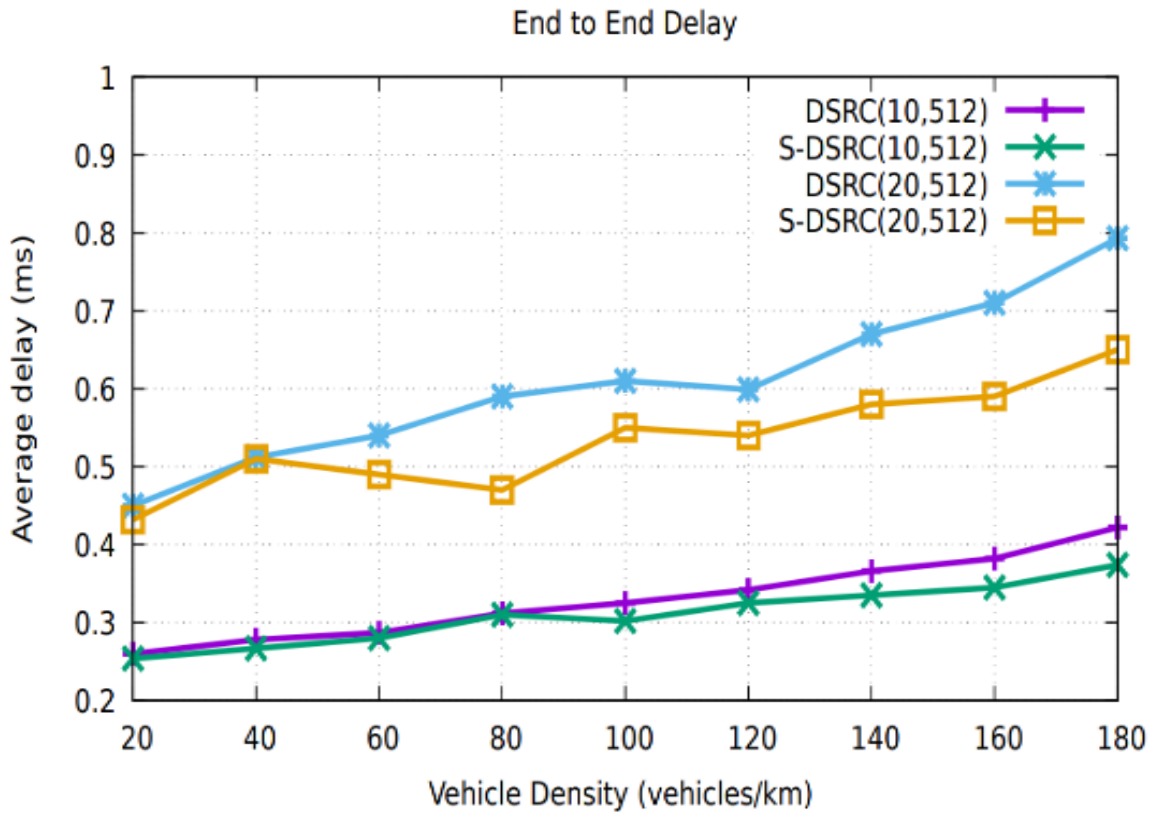

**Fig 7** – Average delay of sending messages between the vehicles

Fig 7 and Fig 8 shows the graph of end to end delay of the messages between the vehicles and packet delivery ratio of successful attempts between the vehicles. The x axis is with a vehicle density of 20 to 180 vehicles per km. The data rate or the bandwidth is varied by 10 and 20 mbps. The results tells us that the S DSRC is always less performing than the DSRC protocol due to the issue of key establishment and encryption/decryption process. But the purpose of this work is to improve security of the messages transmitted between the vehicles.

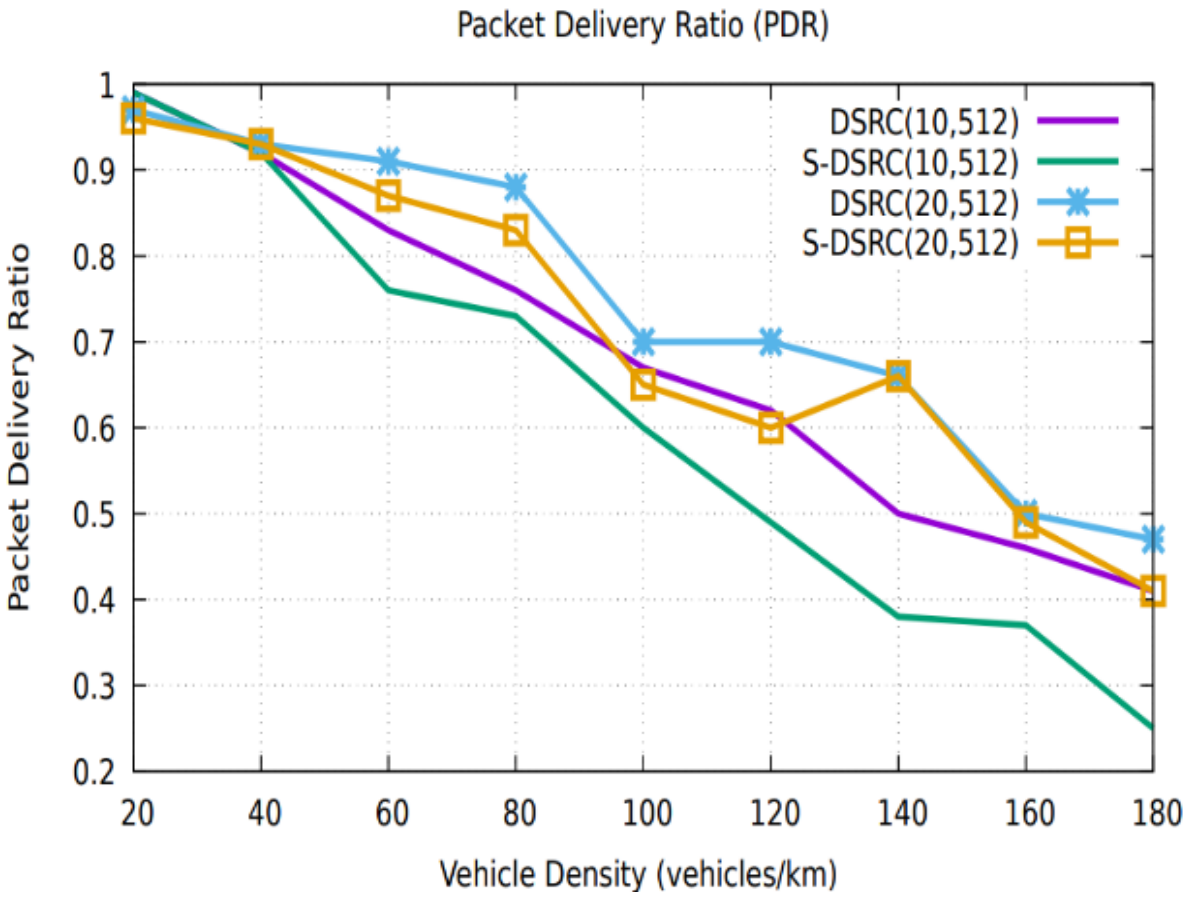

**Fig 8 –** Packet delivery ratio of the network while using Secure DSR

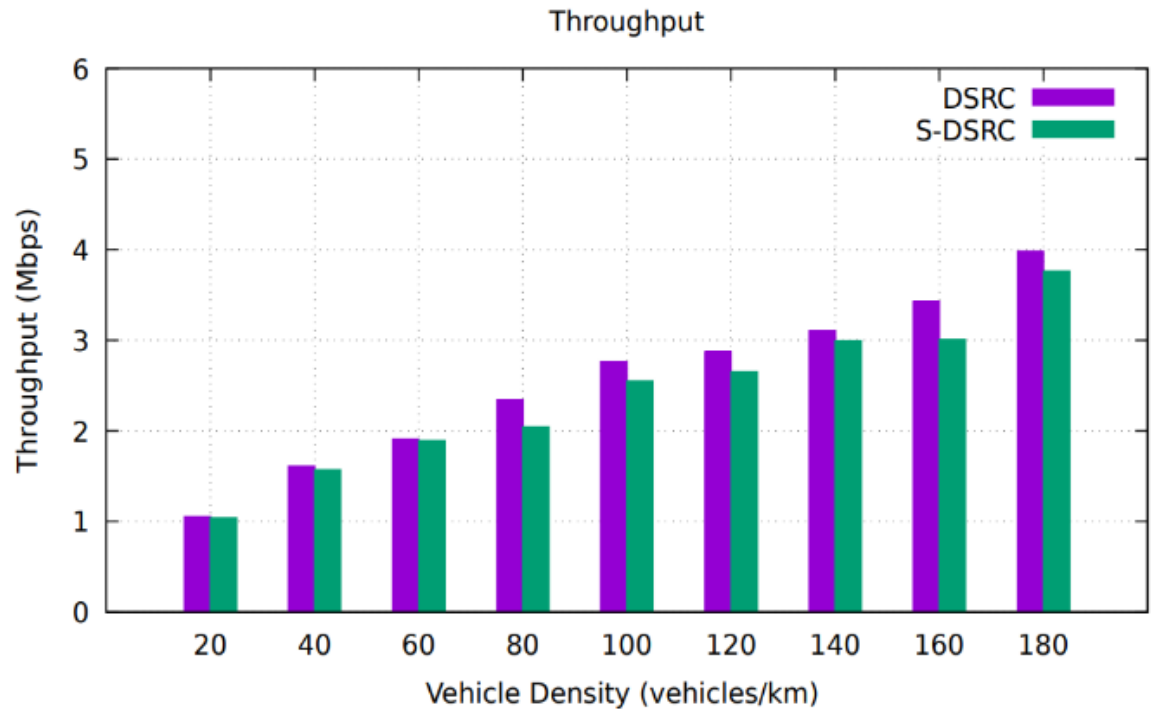

**Fig 9 –** Throughput of the vehicular network

Fig 9 and Fig 10 shows the throughput of the network and packet loss rate of the network. The throughput is decided on various factors like the vehicle speed, data rate, packet arrival rate and the packet size. Since the Secure DSRC protocol establishes a key, there could be some latency due to which the performance is slightly lower than that of plain DSRC protocol. Fig 11 on the other hand is calculated based on the speed of the cars where they are moving at two different speeds 30km/h and 60km/h. We have taken only two speeds as these speed limits are safe in city driving. The packet loss rate is not so deviating even when the vehicles are moving at two different speeds. From the graph, we could see a slight variation in the values and the average values are also in the smaller difference. This indicates that if the vehicles move at a nominal speed of 60km/h, then the packet loss rate is optimal compared to the slow moving of vehicles.

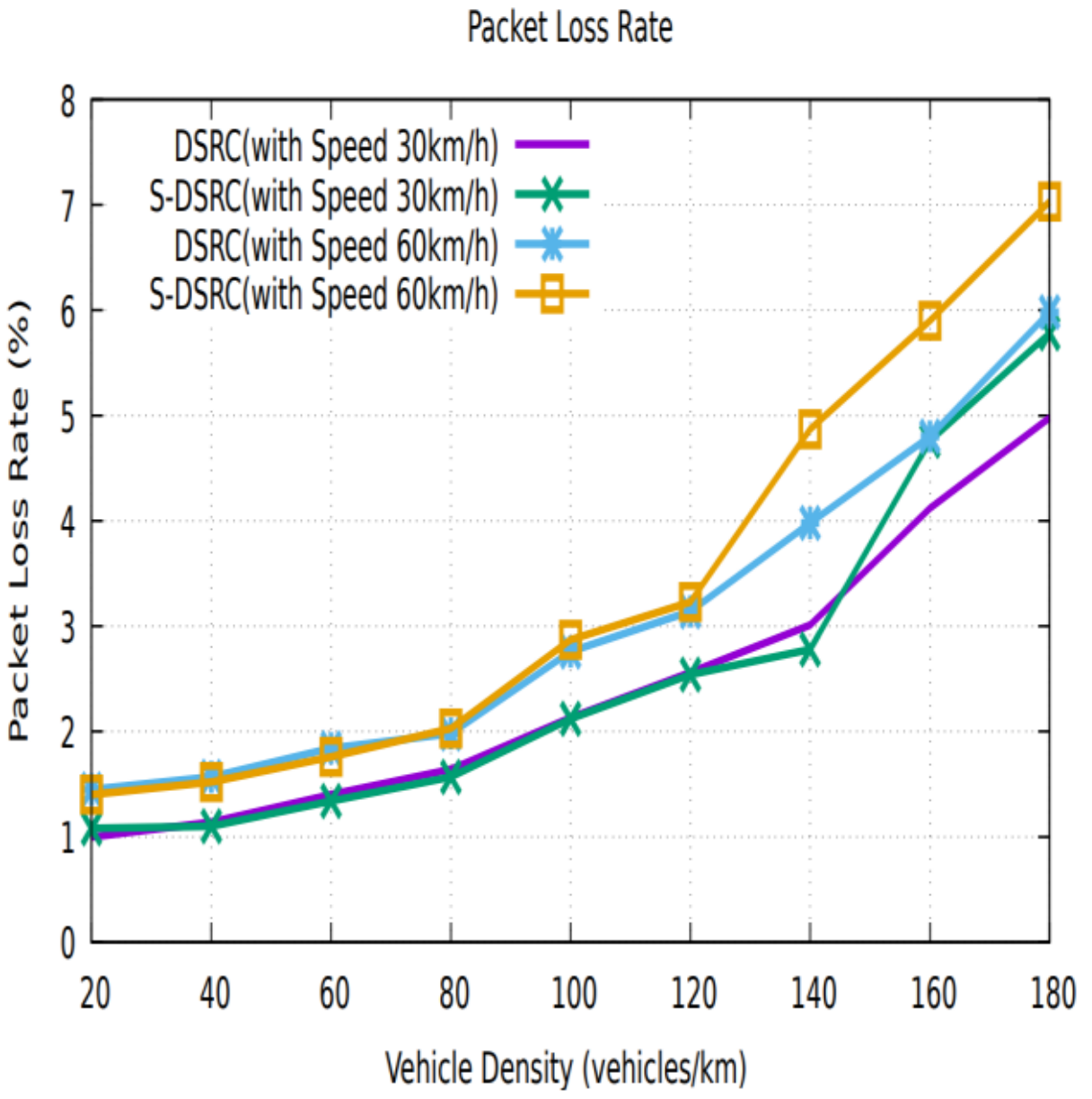

**Fig 10 –** Packet loss rate while using SDSRC and DSRC

Finally, the SDSRC protocol takes some times to establish the keys in a given time. In many security systems, the key generation process will be happening when the nodes are stationery or moving at a lower speed. Here in our work, the key establishment is happening when the vehicles are moving at two different speeds. Hence depends on the speed and mobility of the network, the key establishment time is varying as shown in Fig 11. For a vehicle density of 200 vehicles/km, the key establishment takes almost 14ms which indicates that number of vehicles per km is a key factor for the key establishment process. Also as per the Fig 3, the vehicular data access, road side unit access requests are taken from the cloud provisioning system and hence the latency is included there as well.

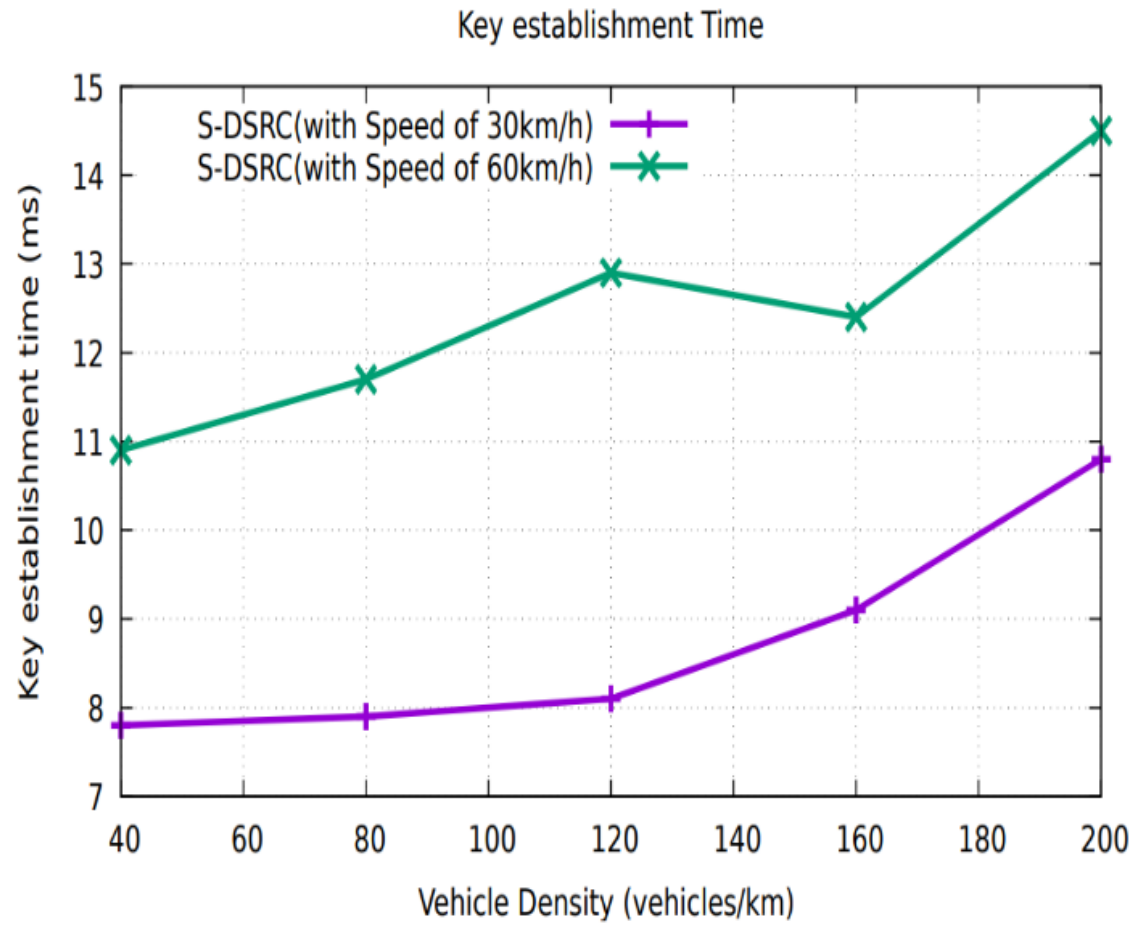

**Fig 11 –** Key establishment time while the vehicles were moving at various speeds

Since the key generation process is occurring between the vehicles, but the access and storage is happening at the cloud. Hence there could be a latency also. That's why we have a higher delay for the same number of vehicles with SDSRC protocol as shown in Fig. 8.

## Conclusion

In this work, we have implemented a secure DSRC protocol with three phases of operation to make the basic safety messages to be secure enough so that they cannot be tampered or hacked into. The setup phase, attributor phase, vehicular data upload phase, vehicular data request phase and finally the road side unit access phase. In all these phases, the data requested is comes from the cloud as most of the date stored directly into the cloud. So, the latency of the system is at stake compared to the non cloud environment. But this system is tested with a non cloud system as well, the same results were appeared when the system is testing against the DSRC protocol. But SDSRC behaves differently in the cloud and the non cloud environment which is the study taken in this work. In this work, we have had many assumptions like speed of the vehicle to be at 60km/h which is a nominal speed requirement of India, but the same work can be extended to other parts of the world for higher speeds and suitably the performance metrics can be measured for that. Also in this work, only cars were taken for simulation in SUMO, but we can include other vehicles like trucks, buses, pedestrian, cycles, etc. Those will be the scope of our future work.